\documentclass[twocolumn,aps,showpacs,amsmath,amssymb,floatfix,superscriptaddress]{revtex4}
\usepackage{graphicx}
\usepackage{dcolumn}
\usepackage{bm}
\usepackage{array}
\usepackage{float}
\usepackage{supertabular}
\usepackage{longtable}
\begin{document}
\title{Citation Statistics From More Than a Century of Physical Review}
\author{S.~Redner\footnote{Address for 2004-05: Theory Division 
and Center for Nonlinear Studies, Los Alamos National Laboratory, Los
Alamos, New Mexico 87545 USA}}
\email{redner@bu.edu}
\affiliation{Center for BioDynamics, Center for Polymer Studies, 
and Department of Physics, Boston University, Boston, MA, 02215}

\begin{abstract}
  
  We study the statistics of citations from all Physical Review journals for
  the 110-year period 1893 until 2003.  We discuss basic properties of the
  citation distribution and find that the growth of citations is consistent
  with linear preferential attachment.  We also investigate how citations
  evolve with time.  There is a positive correlation between the number of
  citations to a paper and the average age of citations.  Citations {\it
  from} a publication have an exponentially decaying age distribution; that
  is, old papers tend to not get cited.  In contrast, the citations {\it to}
  a publication are consistent with a power-law age distribution, with an
  exponent close to $-1$ over a time range of 2 -- 20 years.  We also
  identify one exceptionally strong burst of citations, as well as other
  dramatic features in the time history of citations to individual
  publications.

\end{abstract}


\maketitle

\section{Introduction}

In this article, we study quantitative features of the complete set of
citations for all publications in Physical Review (PR) journals from the
start of the journal in July 1893 until June 2003 \cite{PR}.  This corpus
provides a comprehensive dataset from which we can learn many interesting
statistical facts about scientific citations.  An especially useful aspect of
this data is that it encompasses a continuous span of 110 years, and thus
provides a broad window with which to examine the time evolution of citations
and the citation history of individual publications.

The quantitative study of citations has a long history in bibliometrics, a
subfield of library and information science (see {\it e.g.,} \cite{ER90} for
a general introduction and \cite{G72,S73,SG74} for leads into this
literature).  The first study of citations by scientists was apparently made
by Price \cite{P72,P76}, in which he in which he built upon the original model
of Yule \cite{Y24} and of Simon \cite{S55} to conclude that the distribution
of citations had a power-law form.  It is also worth mentioning much earlier
bibliometric work by Lotka \cite{L26} and by Shockley \cite{S57} on the
distribution of the number of publications by individual scientists.  In the
context of citations, Price termed the mechanism for a power-law citation
distribution as cumulative advantage \cite{M73}, in that that rate at which a
paper gets cited could be expected to be proportional to its current number
of citations.  This mechanism is now known as preferential attachment
\cite{BA99} in the framework of growing network models.

Recently, larger studies of citation statistics were performed that made use
of datasets that became available from the Institute for Scientific
Information (ISI) \cite{ISI} and from the SPIRES database \cite{spires}.
Based on data for top-cited authors, the citation distribution for
individuals was argued to have a stretched exponential form \cite{LS98}.  On
the other hand, by analyzing a dataset of 783,339 papers from all
ISI-cataloged journals and all 24,296 papers in Physical Review D from 1975
until 1994, a power-law citation distribution was inferred \cite{R98}, with
an exponent of $-3$.  This conclusion coincided with subsequent predictions
from idealized network models that grow by preferential attachment
\cite{BA99,KRL00,DMS00,KR01}.  This result was also in accord with the
expectation from Price's original work \cite{P76}.  Finally, it is worth
mentioning a number of very recent data-driven statistical studies of
collaboration and citation networks \cite{N01,B02,LLJ03,BMG04,N04,GMY04},
that are based on a articles from MEDLINE, arXiv.org, NCSTRL, and SPIRES
\cite{N01,LLJ03}, mathematics and neuroscience articles \cite{B02}, and the
Proceedings of the National Academy of Sciences \cite{BMG04}.

The present work is focused on the citation statistics of individual articles
in all PR journals.  While the total number of citations contained in our
study is less than half of what was previously considered in Ref.~\cite{R98}
(approximately 3.1 million vs.\ 6.7 million), the new data encompasses 110
years of citations from what is arguably the most prominent set of archival
physics journals after 1945.  Thus we are able to uncover a variety of new
features associated with the time history of citations, such as highly
correlated bursts of citations, well-defined trends, and, conversely,
downturns in research activity.

It is important to appreciate, however, that citation data from a single
journal, even one as central as Physical Review, has significant omissions.
As we shall discuss in the concluding section, the ratio of the number of
internal citations (cites to a PR papers by other PR publications; this
dataset) to total citations (cites to a PR paper by all publications) is as
small as 1/5 for well-cited elementary-particle physics publications.  There
are also famous papers that did not appear in PR journals, as well as
highly-cited authors that typically did not publish in PR journals.  This
tension between PR and non-PR journals is influenced by global socioeconomic
factors, as well as by more recent opportunistic considerations, such as page
charge policies, and the creation of electronic archives and electronic
journals.  All these factors serve to caution the reader that the
observations presented here are only a partial glimpse into the true citation
impact of physics research publications.

\section{Citation Data}
\label{cite-data}

\subsection{General Facts}

The data provided by the Physical Review Editorial Office covers the period
1893 (the start of the journal) through June 30, 2003.  A sample of the data
is given below (with PRB = Phys.\ Rev. B, PRE = Phys.\ Rev.\ E, PRL = Phys.\
Rev.\ Lett., RMP = Rev.\ Mod.\ Phys., etc.)  \cite{abbrev}: \medskip

{\small
\indent\indent\indent   PRB  19 1203 1979 $|$ PRB  20 4044 1979\hfil\break
\indent\indent\indent   PRB  19 1203 1979 $|$ PRB  22 1096 1980\hfil\break
\indent\indent\indent   PRB  19 1213 1979 $|$ PRB  27 380 1983\hfil\break
\indent\indent\indent   PRB  19 1225 1979 $|$ PRB  24 714 1981\hfil\break
\indent\indent\indent   PRB  19 1225 1979 $|$ PRL  55 2991 1985\hfil\break
\indent\indent\indent   PRB  19 1225 1979 $|$ PRB  38 3075 1988\hfil\break
\indent\indent\indent   PRB  19 1225 1979 $|$ RMP  63 63 1991\hfil\break
\indent\indent\indent   PRB  19 1225 1979 $|$ PRE  62 6989 2000\hfil\break
}

To the left of the vertical line is the {\it cited\/} paper and to the right
is the {\it citing} paper.  In the above sample, Phys.\ Rev.\ B {\bf 19},
1225 (1979) was cited 5 times, once each in 1981, 1985, 1988, 1991, and 2000.
There are 3,110,839 citations in the original data set and the number of
distinct publications that have at least one citation is 329,847.  Since
there are a total of 353,268 publications \cite{doyle}, only 6.6\% of all PR
publications are uncited; this is much smaller than the 47\% fraction of
uncited papers in the ISI dataset \cite{R98}.  The average number of
citations for all PR publications is 8.806.  We emphasize again that this
dataset does not include citations to or from papers outside of PR journals.

There are a variety of amusing basic facts about this citation data.  The
329,847 publications with at least 1 citation may be broken down as follows:

\medskip
{\small\begin{tabular}{rrlcr}

    ~ &  11 &publications with &$>$& 1000 citations\\
   ~  &  79 &publications with &$>$&  500 citations\\
  ~  &  237 &publications with &$>$&  300 citations\\
 ~  &  2,340 &publications with &$>$&  100 citations\\
 ~  &  8,073 &publications with &$>$&   50 citations\\
 ~ & 245,459 &publications with &$<$&   10 citations\\
 ~ & 178,019 &publications with &$<$&    5 citations\\
 ~ &  84,144 &publications with & &    1 citation\\
\end{tabular}
}

\smallskip
\noindent Not unexpectedly, most PR papers are negligibly cited.

For studying the time history of citations, we define the age of a citation
as the difference in the year that a citation occurred and the publication
year of the cited paper.  For all PR publications, the average citation age
in 6.2 years.  On the other hand, for papers with more than 100 citations,
the average citation age is 11.7 years.  The average age climbs to 14.6 years
for publications with more than 300 citations and 18.9 years for the 11
publications with more than 1000 citations.  As one might anticipate,
highly-cited papers are long-lived.  Conversely, papers with only young
citations, are meagerly cited in general.  For example, for publications
(before 2000) in which the average citation age is less than 2 years, the
average number of citations is 3.55.

\subsection{Accuracy}

Because individual authors typically generate citations, a natural concern is
their accuracy.  In recent years, cross-checking has been instituted by the
Physical Review Editorial Office to promote accuracy.  In older papers,
however, a variety of citation errors exist \cite{S02}.  One can get a sense
of their magnitude by looking at the reference lists of old PR papers in the
online PR journals (prola.aps.org).  References to PR papers that are not
hyperlinked typically are erroneous (exceptions are citations to proceedings
of old APS meetings, where the page of the cited article generally does not
match the hyperlinked first page of the proceedings section).  By scanning
through a representative set of publications, one will see that such
citations are occasional but not rare.

While author-generated citation errors, caused by carelessness or propagation
of erroneous citations, are hard to detect systematically, the following
types of errors are easily determined:

\begin{enumerate}
  
\item Old citations are potentially suspect.  A typical example occurs when
  author(s) meant to cite, for example, Phys.\ Rev.\ B {\bf 2}, xxx (1970),
  but instead cited Phys.\ Rev.\ {\bf 2}, xxx (1913).  There are 14807
  citations older than 50 years in the initial dataset, with 4734 of these to
  a publication with a single citation, a feature that suggests an erroneous
  citation.  By looking every hundredth of these 4734 citations, 39 out of 47
  ($\approx 83\%$) were in fact incorrect.  The accuracy rate improves to
  approximately 50\% for the 606 papers with 2 citations and presumably
  becomes progressively more accurate for publications with a larger number
  of citations.
  
\item Citations to pages beyond the total number of pages in the cited volume
  (partially overlaps with item 1).  In vols.\ \mbox{1--80} of Phys.\ Rev.\
  (until 1950), there are 4152 such errors out of 125,240 citations.  In
  vols.\ 1--85 of PRL (which use conventional page numbers), there are 2777
  beyond-page errors out of 912,394 citations, with more recent citations
  being progressively more accurate.  For example, there are only 11
  beyond-page errors in vols.\ 80--85, while there are 145 such errors in
  vol.\ 23 alone.
  
\item Acausal citations; that is, a citation to a publication in the future.
  There are 1102 such errors.
  
\item Truncated page numbers.  After 2001, PR papers are identified by a
  six-digit number that begins with a leading ``0'', rather than a
  conventional page number.  This leading digit was not included 438 times.
  
\item Page numbers in vols.\ 133--139 of Phys.\ Rev.\ were prepended by an A
  or B, a convention that fostered citation errors (see Appendix.~\ref{most} on
  most-cited papers).
  
\item Two papers were referred to at once, {\it e.g.}, PRL {\bf 33}, 100,
  300, (1990) when the lazy authors should have cited PRL {\bf 33}, 100,
  (1990) and PRL {\bf 33}, 300 (1990).
 
\end{enumerate}

Additionally, there were easily-correctable mechanical defects in the
original data.  These include:

\begin{enumerate}
  
\item In volumes with more than 10,000 pages, the comma in the page number
  sometimes appeared as a non-standard character.  The number of such errors
  was approximately 10,000.
  
\item Some lines contained either html or related markup language, or other
  unusual characters.
  
\item Annotated page numbers.  For example, citations to the same paper could
  appear as PRE {\bf 10}, 100 (1990) and PRE {\bf 10}, 100(R) (1990).
  Annotations included (a), A, (A), (b), B, (B), (BR), (e), E, (E), L, (L),
  R, (R), [R], (T), (3), (5).  Some have clear meanings (letter, erratum,
  (A), A, and (a) for meeting notes in the early APS years).  The number of
  such lines was approximately 1500.
  
\end{enumerate}

In summary, the number of readily-identifiable citation errors is of the
order of 10,000, an error rate of approximately $\frac{1}{3}\%$.  The number
of non-obvious errors, {\it i.e.}, citations where the volume, page number,
and publication year are not manifestly wrong, is likely much higher.
However, upon perusal of subsets of the data, the total error rate appears to
be of the order of a few percent, and is considerably smaller in recent
years, even as the overall publication rate has increased \cite{error}.  With
these caveats, we now analyze the citation data.

\section{The Citation Distribution}
\label{sec-cite-dist}

\begin{figure}[ht] 
 \vspace*{0.cm}
 \includegraphics*[width=0.45\textwidth]{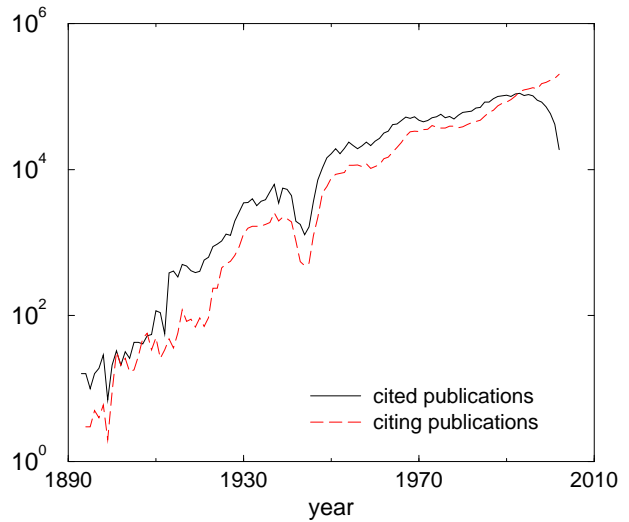}
\caption{Total number of PR citations by year.  
\label{total-by-year}}
\end{figure}

One basic aspect of citations is their rapid growth in time
(Fig.~\ref{total-by-year}), a feature that mirrors the growth of PR journals
themselves.  This growth needs to be accounted for in any realistic modeling
of the distribution of scientific citations (see {\it e.g.}, \cite{DM01}).
The number of citations by {\it citing} papers published in a given year is
shown as the dashed curve, and the number of citations to {\it cited\/}
papers that were published in a given year is shown as the solid curve.
Notice the significant drop in citations during World War II, a feature that
was also noted by Price \cite{P72}.  The fact that the two curves are so
closely correlated throughout most of the history indicates that most
citations are to very recent papers.  Another noteworthy feature is that the
long-term growth rate of citations is smaller after WWII than before.  The
very recent decay in cited publications occurs because such papers have not
yet sufficient time to be completely cited.  Finally, note that area under
the two curves must be the same.

\begin{figure}[ht] 
 \vspace*{0.cm}
 \includegraphics*[width=0.45\textwidth]{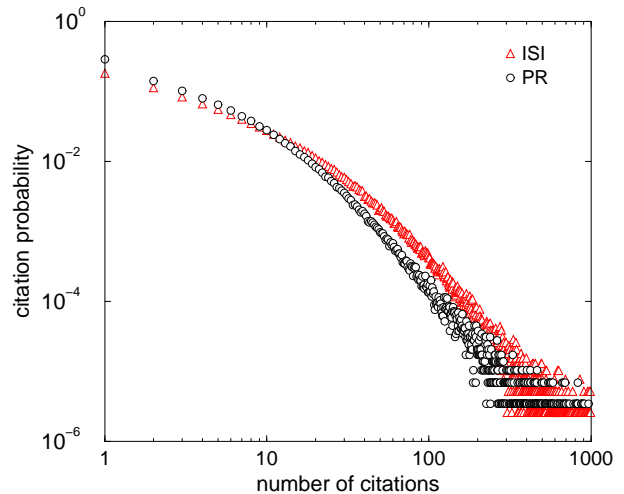}
\caption{Normalized citation distributions for all papers in Phys.\ Rev.\
  from 1893 to 2003 ($\circ$) and from the ISI dataset ($\Delta$) consisting
  of scientific papers published in 1981 that were cited between 1981 and
  1997 (from Ref.~\cite{R98}).
\label{total-cites}}
\end{figure}

We next show the citation distribution for the entire dataset
(Fig.~\ref{total-cites}).  This distribution is visually similar to that in
Ref.~\cite{R98} for the corresponding ISI distribution.  While there is
systematic curvature in the data on a double logarithmic scale, a Zipf plot
of the ISI data, which focuses on the large-citation tail, suggested a
power-law form for the citation distribution.  A similar conclusion for the
citation distribution can thus be expected for the PR data.  A
straightforward power-law fit to the data in the range of 50 -- 300 citations
gives an exponent of $-2.55$ for both the PR and ISI data.  However, as
argued in Ref.~\cite{R98} by using a Zipf plot to account for publications
whose citation histories are not yet complete, the exponent of the ISI
citation distribution is consistent with the value $-3$.

\begin{figure}[ht] 
 \vspace*{0.cm}
 \includegraphics*[width=0.45\textwidth]{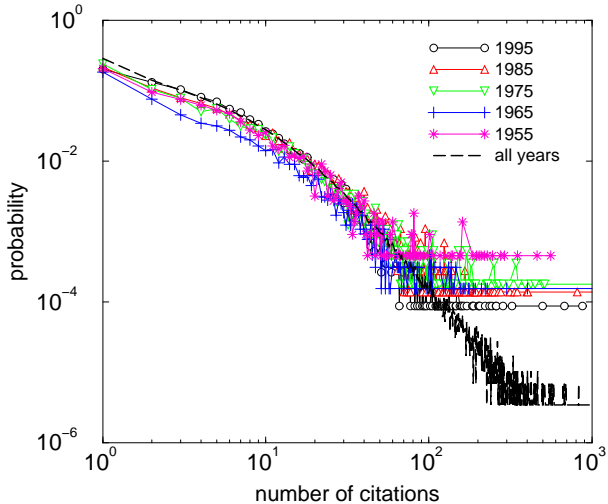}
\caption{Normalized citation distributions in selected years.
\label{cites-by-year}}
\end{figure}

To check if the nature of the citation distribution is affected by the growth
of PR journals, we plot in Fig.~\ref{cites-by-year} the citation distribution
to papers published in selected years, along with the total citation
distribution.  We see that these yearly distributions closely match the total
distribution except at the large-citation tail.  There is a hint that the
small-citation tail of the distribution ($\alt 20$) is qualitatively
different than the rest of the distribution.  A natural suspicion is that
self-citations might play a significant role because papers with few
citations are likely to be predominantly self-cited.

\section{The Attachment Rate}

An important theoretical insight into how scale-free networks develop was the
realization that the rate at which a new node attaches to a
previously-existing node is an increasing function of the degree of the
target node \cite{BA99}; this is the mechanism of preferential attachment.
Here, the degree of a node is the number of links that are attached to the
node, or equivalently, the number of citations to a publication.  More
precisely, the network growth is controlled by the rate $A_k$ at which a new
node attaches to a previously-existing node of degree $k$.  In the context of
citations, $A_k$ then gives the rate at which an existing paper with $k$
citations currently gets cited.  In this section, we study the attachment
rate for all PR publications.

Earlier studies of this attachment rate \cite{B02,JNB01} examined citation
data for 2 years of PRL, both a mathematic and a neuroscience co-authorship
network over 8 years, a century of data of a network of actor that
co-appeared in all movies, and the Internet from 1997-2001.  For the citation
network of PRL and the Internet, linear preferential attachment was observed,
while the attachment rate grew slower than linearly with $k$ for the other
examples.  As was first found in \cite{BA99}, an attachment rate that is
linearly proportional to $k$, $A_k \propto k$ leads to a power-law node
degree distribution, in which the number of nodes of degree $k$, $n_k$,
scales as $n_k\sim k^{-\nu}$.  In the specific case where $A_k=k$, the
exponent $\nu=3$ \cite{BA99,KRL00,DMS00}, while for attachment rates that are
asymptotically linear in $k$, the degree distribution exponent can be tuned
to have {\it any} value in the range $(2,\infty)$ \cite{KRL00,DMS00,KR01}.

\begin{figure}[ht] 
 \vspace*{0.cm}
 \includegraphics*[width=0.45\textwidth]{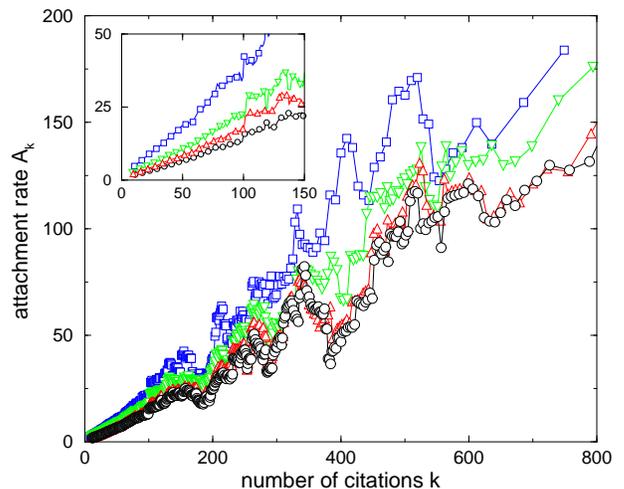}
\caption{The attachment rate $A_k$ for the PR citation data.  Shown are the
results when the initial network is based on citations from 1990-1999
($\square$), 1980-1999 ($\bigtriangledown$), 1970-1999 ($\bigtriangleup$),
and 1892-1999 ($\circ$).  The data has been averaged over 5\% of the total
number of data points for each case.  The inset shows the rate for the range
of $\leq 150$ citations.
\label{Ak}}
\end{figure}

For our analysis of the attachment rate, we first construct the degree
distribution of the initial network by taking a specified time window of the
initial PR data.  We then specify a second time window during which the
attachment rate is measured.  The first window ran from a specified starting
year (see Fig.~\ref{Ak}) until 1999, while the second window was the year
2000.  Operationally, we counted the number of times each paper was cited
from the starting year until 1999 (giving $k$) and then counted how many
times each such paper was then cited in 2000.  From this data, $A_k$ is
proportional to the number of times a paper with $k$ earlier citations was
cited in 2000.  As shown in Fig.~\ref{Ak}, the qualitative results do not
depend strongly on the starting year.  This weak dependence stems from the
fact that the total publication rate in Physical Review has grown so rapidly
that the bulk of the publications and citations are in the last decade.  Thus
there is relatively little change when data from earlier decades are
included.

More importantly, the data suggests that the attachment rate $A_k$ is a
linear function of $k$, especially in the range of $\leq 100$ citations.
After dividing out the irrelevant proportionality constant in $A_k$, that is,
$A_k=ak+b\to k+b/a\equiv k+\lambda$, we find that the initial attractiveness
parameter $\lambda$ is in the range $(-0.28,+0.37)$ for the data shown in the
figure.  When the raw data for $A_k$ is averaged over a 5\% range, there is a
systematic tendency for $\lambda$ to decrease to slightly negative values.
If one follows the analysis method of \cite{JNB01} and considers the
integrated rate $I_k\equiv \int^k A_{k'}\, dk'$, then a double logarithmic
plot of $I_k$ versus $k$ gives a much smoother, nearly straight curve.  By
fitting this curve to a straight line, we obtain a slope in the range 1.91 --
2.05, depending on the subrange of the data over which the fit is made.

Overall, the results strongly suggest that the attachment rate $A_k$ has the
linear preferential form $k+\lambda$, with a small value for the initial
attractiveness parameter $\lambda$.

\section{Age Characteristics of Citations}

One of the more useful aspects of having 110 years of citation data is the
ability to study their age structure.  In theoretical modeling of growing
networks, it was found that the joint age-degree distribution of the nodes in
a network provides many useful insights about network structure
\cite{KR01,extreme}.  

\begin{figure}[ht] 
 \vspace*{0.cm}
 \includegraphics*[width=0.45\textwidth]{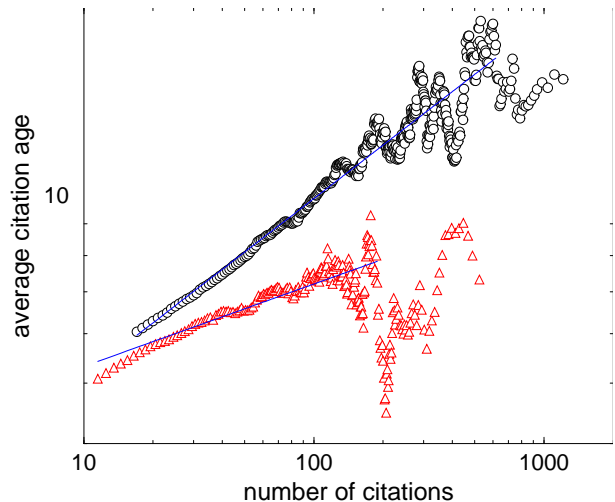}
\caption{Average age of citations to a given paper versus number of citations
  to this paper.  Shown are all publications ($\circ$) and for dead
  publications ($\Delta$).  Both sets of data were averaged over a 5\% range.
  Straight lines are the best fits for the range shown.
\label{av-age}}
\end{figure}

Empirically, unpopular papers are typically cited only soon after publication
(if at all) and then disappear.  We thus expect that the number of citations
to a paper and the average age of these citations are positively correlated.
Fig.~\ref{av-age} shows this average citation age versus total number of
citations.  It is also revealing to distinguish between ``dead'' and
``alive'' papers in this plot.  While it is not possible to be definitive, we
define dead papers as those with less than 50 citations and where the average
age of its citations is less than one-third the age of the paper itself.
Similar definitions of publication death and its ramifications on citation
networks have been studied in Ref.~\cite{P00,LJL04,HS04}.  Based on examining the
actual PR citation data, our definition of dead papers appears generous, as
almost all publications that are considered as dead by our criterion remain
dead.

As expected, there is a positive correlation between the average citation age
$\langle A\rangle$ and the number of times $N$ that a publication has been
cited.  The dependence is very systematic for fewer than 100 citations, but
then fluctuates strongly beyond this point.  Over the more systematic portion
of the data, power-law fits suggest that $\langle A\rangle \sim N^\alpha$,
with $\alpha\approx 0.285$ for all publications and $\alpha\approx 0.132$ for
dead publications.  It was found previously that the average number of
citations is positively correlated to the age of the publication itself in
idealized growing networks \cite{KR01}, and it would be worthwhile to
determine whether a similar positive correlation extends to the average
citation age.

\begin{figure}[ht] 
 \vspace*{0.cm}
 \includegraphics*[width=0.45\textwidth]{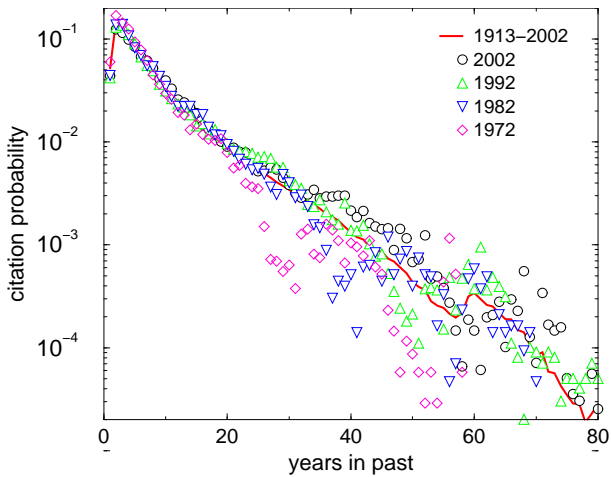}
\caption{The distribution of the ages of citations contained in the reference
  lists of publications that were published in selected years.  Also shown is
  this same citing age distribution for the period 1913-2002.
\label{citing-memory}}
\end{figure}

\begin{figure}[ht] 
 \vspace*{0.cm}
 \includegraphics*[width=0.45\textwidth]{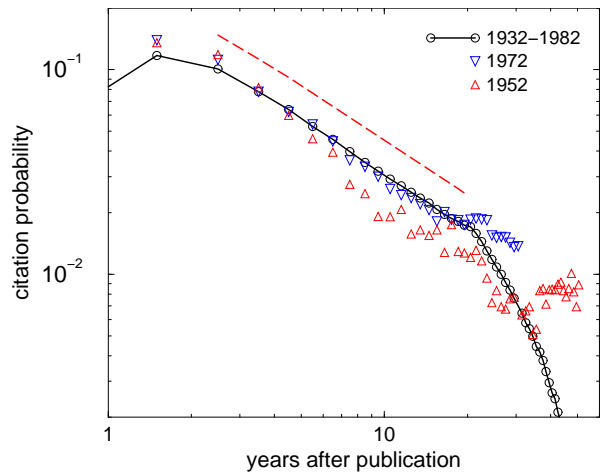}
\caption{Distribution of the ages of citations to cited papers in
  selected years, as well as the integrated data over the period 1932-1982.
  The dashed line is the best fit to the data in the range 2 -- 20 years
  (displaced for visibility).
\label{cited-aggregate}}
\end{figure}

A more basic quantity than the average citation age is the underlying age
distribution of citations.  As previously alluded to in
Sec.~\ref{sec-cite-dist}, there are two distinct distributions of citation
ages.  One is the distribution of citation ages from {\it citing\/}
publications (Fig.~\ref{citing-memory}).  This refers to the age (years in
the past) of each citation in the reference list of a given paper.  The
second, and more fundamental, distribution refers to the ages of citations to
{\it cited\/} publications (Fig.~\ref{cited-aggregate}).  For example, for a
paper published in 1980 that is subsequently cited once in 1982, twice in
1988 and three times in 1991, the (cited) citation age distribution has
discrete peaks at 2, 8 and 11 years, with respectively weights 1/6, 1/3, and
1/2.

The distribution of citing ages is shown in Fig.~\ref{citing-memory} for
papers published in selected years, as well as the distribution integrated
over the years 1913-2002.  In the range of 2 -- 20 years, the distribution
decays exponentially in time, with a 10-fold decrease in citation probability
across a 13-year time span.  For longer times, there is a slower exponential
memory decay of approximately 23 years for each decade drop in citation
probability.  However, this decay is masked by the influence of WWII.  For
example for the 1972 data, there is a pronounced dip between 25 -- 30 years.
This dip moves 10 years earlier for each 10-year increase in the publication
year.  If this dip was not present, it does appear that the older citation
data would exhibit data collapse.  Finally, notice that the integrated
distribution has a perceptible WWII-induced dip at 57 years in the past,
indicative of the fact that most PR publications have appeared in the last
decade.

Since the number of old publications is a small fraction of all publications,
the integrated distribution does not change perceptibly whether or not very
early publications are included.  The annual and the integrated distributions
are all quite similar and show the same range of memory for old papers,
independent of their publication year.  Thus the forgetting of old papers is
a primary driving mechanism for citations.  Previous studies by Price
\cite{P72} (based on a smaller dataset) and subsequently by various authors
\cite{P00,LJL04,HS04} have considered the role of decaying memory on the
citation network.  Theoretical models of decaying memory on the structure of
growing networks were studied in \cite{DM00,KE02}.  According to
Ref.~\cite{DM00}, exponentially decaying memory leads to an exponential
degree distribution in a network that grows at a constant rate.  We argue
that the PR citation network still has a power-law degree distribution
because of the exponential growth in the number of publications with time, an
effect that reduces the importance of finite citation memory.

Notice also that the idealized preferential attachment mechanism gives
preferential citations to older papers -- opposite to what is observed.  This
discrepancy again stems from the fact that the total PR citation network has
been growing exponentially with time rather than linearly.  As a result, most
PR papers have been published in the past decade and there are relatively few
old publications available to cite.  Thus while the adage ``nobody cites old
papers anymore'' is figuratively true for PR publications, it is mostly due
to the rapid growth of the journal.

We next present the distribution of cited ages.  Since
Fig.~\ref{citing-memory} suggests that citations to papers younger than 20
years are incomplete, we consider only the cited age distribution of older
publications.  In fact, the cited age distribution for more recent
publications has a sharp cutoff when the current year is reached because such
papers are still being cited at a significant rate.

Fig.~\ref{cited-aggregate} shows representative age distributions for papers
published in 1952 and 1972, and as well as the integrated 1932-1982
distribution.  Once again, the integrated distribution does not change
perceptibly if pre-1932 data are included.  In the figure, we add 0.5 to all
ages, so that a citation occurring in the same year as the original paper is
assigned an age of 0.5.  Over the limited range of 2 -- 20 years, the
integrated data is consistent with a power law decay with an associated
exponent of $-0.94$ (dashed line in the figure).  Thus even though authors
tend to have an exponentially-decaying memory in the publications that they
cite, the cited citation age distribution has a much slower (apparently)
power law decay with time.  This is a major result of our data analysis that
should be worthwhile to model.  Once again, the exponential growth of PR
journals and the concomitant increase in the number of citations to past
publications may strongly influence this cited age distribution.

\section{Citation Histories of Individual Publications}

Although we have seen that the collective citation history of all PR
publications is quite systematic, the citation histories of individual
well-cited publications show great diversity and a variety of amusing
features.  A substantial fraction of the citation histories can be roughly
(but not exhaustively) categorized as being either revival of classic works,
major discoveries, or hot publications.  We now present examples from each of
these classes.

\subsection{Revival of Old Classics}

Sometimes a publication will remain long-unrecognized and then suddenly
become in vogue; this has been termed a ``sleeping beauty'' in the
bibliometric literature \cite{R04}.  We (arbitrarily) define this category of
publication as all non-review PR articles (excluding RMP) with more than 300
citations and for which the ratio of average citation age to age of the paper
is greater than 0.75, {\it i.e.}, well-cited papers in which the bulk of
their citations occur closer to the present rather than to the original
publication date.  Remarkably, only 8 papers fit these two criteria.  They
are:

\begin{widetext}

{\small\begin{longtable}{|c|p{0.2in}|>{\hfill}p{0.2in}|>{\hfill}p{0.22in}|
>{\hfill}p{0.275in}|>{\hfill}p{0.25in}|p{3.5in}|p{1.8in}|}
\caption{The 8 PR papers with $>300$ citations and with citation age/paper age $>0.75$.}\label{tab-classic}\\
\hline
\endhead
  Impact   & \multicolumn{4}{c|}{}             & \#~~~&&    \\ 
  Rank    &  \multicolumn{4}{c|}{Publication}  & cites & Title & Author(s) \\ \hline

4&PR& 40& 749& 1932&   568&          On the Quantum Correction for
Thermodynamic Equilibrium & E. Wigner\\ \hline

7&PR& 47& 777& 1935&   532&          
Can Quantum-Mechanical Description of Physical Reality Be Considered Complete?&
\mbox{A. Einstein, B. Podolsky, \&} N.~Rosen\\ \hline

23&PR&  56& 340& 1939&   350&       Forces in Molecules 
& R. P. Feynman\\ \hline  

6&PR& 82& 403& 1951&   678&          Interaction between $d$-Shells in 
Transition Metals.\ II. Ferromagnetic Compounds of Manganese with Perovskite Structure
& C. Zener\\ \hline  

30&PR& 100& 545& 1955&    374&   
Neutron Diffraction Study of the Magnetic Properties of the Series of
Perovskite-Type Compounds [$(1-x$)La,$x$Ca]MnO$_3$ 
&E. O. Wollan \& W. C. Koehler \\ \hline
 
37&PR& 100& 564& 1955&   302&     Theory of the Role of Covalence in 
the Perovskite-Type Manganites [La, M(II)]MnO$_3$   & J. B. Goodenough  \\ \hline

19&PR& 100& 675& 1955&   483&        Considerations on Double Exchange
& P. W. Anderson \& H. Hasegawa\\ \hline 

21&PR& 118& 141& 1960&   519&         Effects of Double Exchange in
Magnetic Crystals &  P.-G. de Gennes\\ \hline 

\end{longtable}
}

\end{widetext}

\begin{figure}[ht] 
 \vspace*{0.cm}
 \includegraphics*[width=0.44\textwidth]{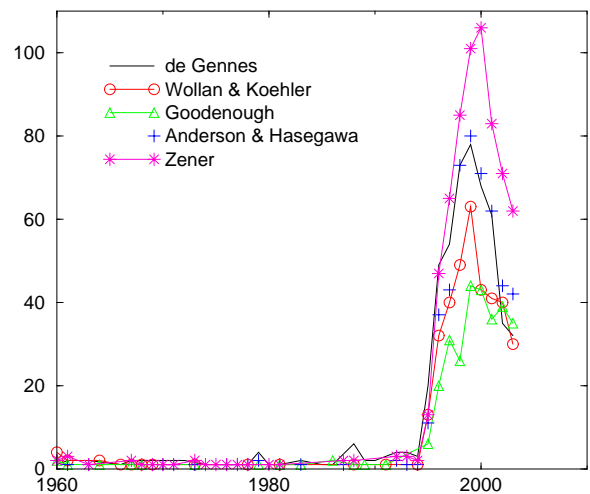}
\caption{Citation histories of the 5 publications of relevance to colossal
  magnetoresistance.
\label{rediscovered}}
\end{figure}

The number of citations in this table have been updated through the end of
2003.  The clustering of citation histories of the last 5 of these 8
publications is particularly striking (Fig.~\ref{rediscovered}).  These
interrelated papers were written between 1951 and 1960, with 3 in the same
issue of Physical Review.  They were all concerned with the ``double
exchange'' mechanism in manganites with a Perovskite structure.  In
particular, Zener's paper had (17,7,9,4) citations in its first 4 decades and
more than 600 citations subsequently!  Double exchange is the interaction
responsible for the phenomenon of colossal magnetoresistance, a topic that
became extremely popular through the 90's due to the confluence of new
synthesis and measurement techniques in thin-film transition-metal oxides,
the sheer magnitude of the effect, and the clever coining of the term
``colossal'' to describe the phenomenon \cite{CMR}.  The simultaneous
extraordinary burst of citations to these articles in a short period close to
the year 2000, more than 40 years after their original publication, is unique
in the entire history of PR journals.

Of the remaining articles, the publications by Wigner and by Einstein et al.\ 
owe their renewed popularity to the upsurge of interest on quantum
information phenomena.  Finally, Feynman's work presented a new (at the time)
method for calculating forces in molecular systems, a technique that has had
wide applicability in understanding interactions between elemental
excitations in many fields of physics.  This paper is particularly noteworthy
because it is cited by papers from all PR journals: PR, PRA, PRB, PRC, PRD,
PRE, PRL, and RMP!

\begin{figure}[ht] 
 \vspace*{0.cm}
 \includegraphics*[width=0.45\textwidth]{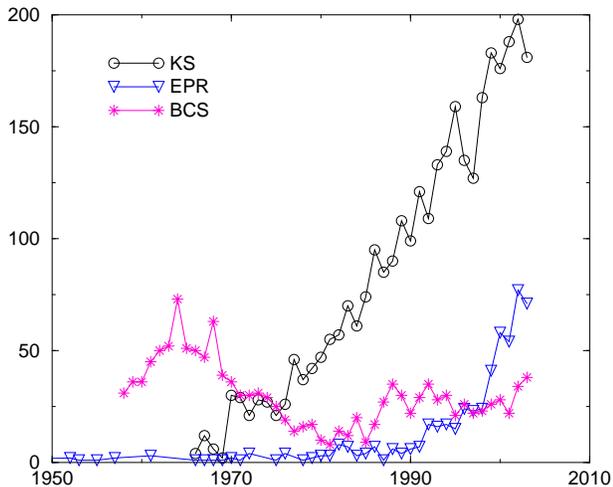}
\caption{Citation history of 3 classic highly-cited publications.  Each is
  identified by author initials (see text).
\label{classic3}}
\end{figure}

Shown also are the citation histories of 3 PR papers that have been famous
over a long time scale (Fig.~\ref{classic3}).  The paper with most citations
in all PR journals is ``Self-Consistent Equations Including Exchange and
Correlation Effects'', Phys.\ Rev.\ {\bf 140}, A1133 (1965) by W.~Kohn \&
L.~J.~Sham (KS), with 3227 citations as of June 2003 (see
Appendix~\ref{most}).  It is amazing that citations to this publication have
been steadily increasing for nearly 40 years.  On the other hand, the paper
``Can Quantum-Mechanical Description of Physical Reality Be Considered
Complete?'', Phys.\ Rev.\ {\bf 47}, 777 (1935) by A.~Einstein, B.~Podolsky,
\& N.~Rosen (EPR) had 82 citations before 1990 and 515 citations subsequently
-- 597 citations in total at the end of 2003.  The longevity of EPR is the
reason for the appearance of this publication on the top-10 citation impact
list in Appendix~\ref{most}.  The current interest in EPR stems from the
revival of work on quantum information phenomena.  Finally, the citation
history of ``Theory of Superconductivity'', Phys.\ Rev.\ {\bf 108}, 1175
(1957) by J.~Bardeen, L.~N.~Cooper, \& J.~R.~Schrieffer (BCS) peaked in the
60's, followed by a steady decay through the mid-80's, with a minimum in the
number of citations in 1985, the year before the discovery of
high-temperature superconductivity.  It is worth emphasizing that BCS is the
earliest PR publication with more than 1000 citations (with 1388 citations at
the end of 2003).

\subsection{Discoveries}

Major discoveries are often characterized by a sharp spike in citations when
the discovery becomes recognized.  We are able to readily detect the subset
of such publications in which a citation spike occurs close to the time of
publication.  We considered all non-review articles (excluding both RMP and
compilations by the Particle Data Group) with more than 500 citations, in
which the ratio of average citation age to age of the publication is less
than 0.4 (Table~\ref{tab-discovery} in Appendix~\ref{hot}).  There are a
total of 11 such publications.  Amusingly, the first 6 of these publications
(1975 and previous) are in elementary-particle or nuclear physics, while the
remaining 5 (1984 and later) are in condensed-matter physics.

\begin{figure}[ht] 
 \vspace*{0.cm}
 \includegraphics*[width=0.45\textwidth]{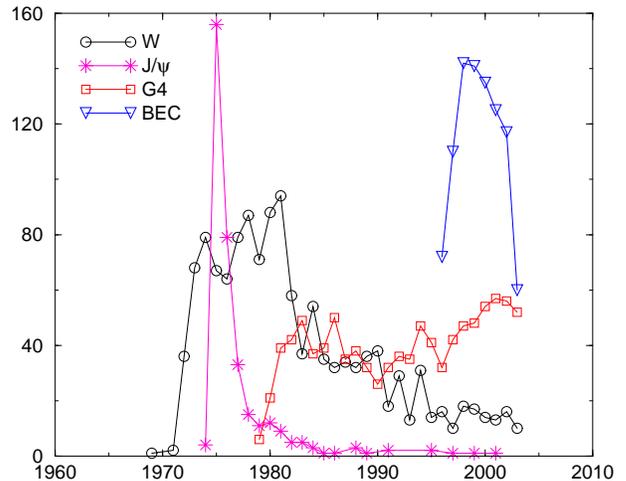}
\caption{Citation history of recent highly-cited discovery publications
  identified by author initials or by topic (see text).
\label{discovery5}}
\end{figure}

By looking more broadly at the citation histories of discovery publications,
we again find considerable diversity (Fig~\ref{discovery5}).  For example,
the average lifetime of citations to the 1974 publications that announced the
discovery of the $J/\psi$ particle -- Phys.\ Rev.\ Lett.\ {\bf 33}, 1404 \&
1406 (1974).  The citation histories of these two publications are
essentially identical and could only be characterized as supernovae.  The
average citation age of these two publications is less than 3 years! The
publication ``Bose-Einstein Condensation in a Gas of Sodium Atoms'', Phys.\
Rev.\ Lett.\ {\bf 75}, 3969 (1995) by K.~B.~Davis, M.-O.~Mewes,
M.~R.~Andrews, N.~J.~van Druten, D.~S.~Durfee, D.~M.~Kurn, \& W.~Ketterle
(BEC) also has a strongly-peaked citation history (but less extreme than the
$J/\psi$ papers), as befits an important discovery in a quickly evolving
field.  Many well-recognized papers that report major discoveries have such a
sharply-peaked citation history.

On the other hand, many discovery papers have a much longer lifetime.  For
example, the paper, ``A Model of Leptons'', Phys.\ Rev.\ Lett.\ {\bf 19},
1264 (1967), by S.~Weinberg, was a major advance in the electroweak theory.
The citation history follows what one might naively anticipate -- a peak, as
befitting a major discovery, but followed by a relatively slow decay (with
1311 citations at the end of 2003).  A very unusual example is ``Scaling
Theory of Localization: Absence of Quantum Diffusion in Two Dimensions'',
Phys.\ Rev.\ Lett.\ {\bf 42}, 673 (1979) by E.~Abrahams, P.~W.~Anderson,
D.~C.~Licciardello, \& T.~V.~Ramakrishnan (the ``gang of four'', G4).  This
paper has been cited between 40 - 60 times nearly every year since
publication, a striking testament to the long-term impact of this publication
to subsequent research.

\subsection{Hot Publications}

\begin{figure}[ht] 
 \vspace*{0.cm}
 \includegraphics*[width=0.45\textwidth]{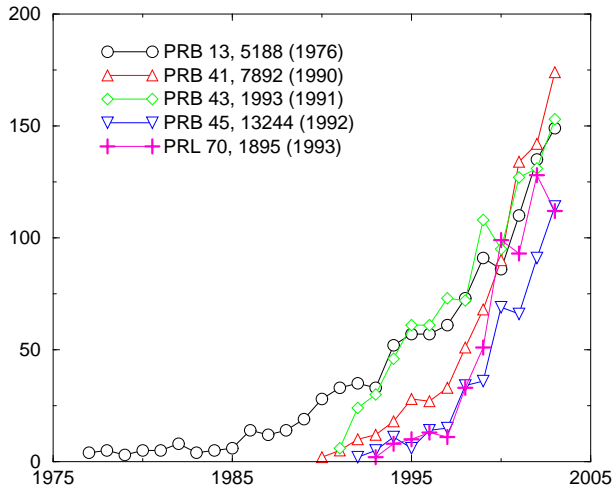}
\caption{Citation histories of 5 recent and highly-cited publications in
  which number of citations are rapidly increasing year by year.
\label{hot5}}
\end{figure}

Finally, it is amusing to identify paper that can be classified as ``hot''.
A listing of the 10 hottest (non-review) PR papers according to the criteria
of $\geq 350$ citations, ratio of average citation age to publication age
greater than 2/3, and citation rate still increasing with time, is given in
Table~\ref{tab-hot} of Appendix~\ref{hot}.  The current citation rate for the
hottest of these articles is unprecedented over the history of Physical
Review (Fig.~\ref{hot5}) and is partially due to the rapid growth in all PR
journals.

A noteworthy aspect of these hot publications is that 3 of them are concerned
with pseudopotential methods, a topic that originates with the seminal
Kohn-Sham article from 1965.  These publications are: ``Soft self-consistent
pseudopotentials in a generalized eigenvalue formalism'', Phys.\ Rev.\ B {\bf
41}, 7892 (1990), by D. Vanderbilt, ``Accurate and simple analytic
representation of the electron-gas correlation energy'', Phys.\ Rev.\ B {\bf
45}, 13244 (1992), by J. P. Perdew and Y. Wang, and ``Efficient
Pseudopotentials for Plane-Wave Calculations'', Phys.\ Rev.\ B {\bf 43}, 1993
(1991) by N.~Troullier \& J.~L.~Martins.  The publication ``Special points
for Brillouin-zone integrations'', by H. J. Monkhorst and J. D. Pack Phys.\
Rev.\ B {\bf 13}, 5188 (1976) develops a generally useful technique for band
structure calculations.  Finally the last publication ``Teleporting an
unknown quantum state via dual classical and Einstein-Podolsky-Rosen
channels'', Phys.\ Rev.\ Lett.\ {\bf 70}, 1895 (1993), C. H. Bennett, et al.\
is concerned with quantum information theory; this is the other major
research area that can be classified as hot, according to recent PR citation
rates to articles in this field.

As a final note, it is amazing that the most-cited 1965 paper by Kohn \& Sham
and the second most-cited 1964 paper by Hohenberg \& Kohn can still be
characterized as hot.  Another striking example of such a citation pattern is
Anderson's 1958 publication on localization in disordered systems, where the
citation rate has had a similar growth as the two previously-mentioned
articles.

\section{Discussion}

The availability of a large continuous body of citation data from a major
physics journal, Physical Review (PR), provides a unique window to observe
how subfields evolve and how individual publications can influence subsequent
research.  It is important to be aware of two basic limitations of citation
data from PR journals only: (i) first, a variety of important physics
articles were not published in PR journals, and (ii) our data does not
include citations to PR articles from articles published in non-PR journals.
The second effect is significant.  By looking at top-cited
elementary-particle physics PR articles in the SPIRES database \cite{top}, we
found that the ratio of internal citations (citations from other PR
publications) to total citations is in range 0.19 -- 0.36.  This ratio
provides a sense of the incompleteness of the PR data.  It would be
worthwhile to extend this citation study to a broader range of physics
journals to see if new citation patterns might emerge.  Finally, as pointed
out in \cite{A02}, citations should not be use guide policy decisions as they
are an imperfect measure of scientific quality.

One of our basic observations is that the development of citations appears to
be described by linear preferential attachment, but with the proviso that
citation memory extends back no more than approximately 20 years.  Much of
this short memory can be ascribed to the exponential growth of PR journals, a
feature that would necessarily favor citations to recent publications.  The
average citation history is very well defined and the age distribution of
citations to a paper is described by a slow power-lay decay up to
approximately 20 years.

At a more descriptive level, we found several striking confluences of
citation activity during the history of Physical Review.  The most prominent
of these -- on the topic of colossal magnetoresistance -- is quite recent,
but are based on work of more than a half century ago.  There are also a
small number of ``hot'' publications that are currently being cited at a
remarkable rate.  Much of this activity revolves around density functional
theory, pseudopotential methods, and the development of accurate techniques
for band structure calculations.  The origin of much of this work is, in
turn, the pioneering Kohn-Sham paper of 1965.  The other clearly-identifiable
topical coincidence in highly-cited publications occurs in quantum
information theory.  Part of the reason for the large citation rate to these
hot papers could well be the larger number of researchers compared to several
decades ago, as well as the rapid availability of preprints through
electronic archives.  Nevertheless, the rapid and recent growth in citations
of these publications seem to portend scientific advances.  Finally, it is
worth noting the large role that a relatively small number of individual
physicists have played in PR publications, with two individuals (Phil
Anderson and Walter Kohn) each co-authoring five articles in the top-100
citation impact list \cite{Re04}.

\acknowledgments{

I am grateful to Martin Blume for giving permission to access the citation
data and for the helpful assistance of Mark Doyle and Gerry Young of the APS
editorial office for providing this data.  I thank Jon Kleinberg for initial
collaboration and many helpful suggestions, Claudia Bondila and Guoan Hu for
writing perl scripts for processing some of the data, and Antonio
Castro-Neto, Andy Cohen, Andy Millis, Claudio Rebbi, and Anders Sandvik for
helpful literature advice.  I also thank K. B\"orner, P. Gangier, P. H\"anggi
and O. Toader for helpful suggestions on the first version of the manuscript.
Finally, I am grateful for financial support by NSF grant DMR0227670 (BU) and
DOE grant W-7405-ENG-36 (LANL).}

\newpage

\begin{widetext}

\appendix

\section{Citation and Impact Rankings}
\label{most}

For general interest, the top-10 cited PR papers, as of June 2003, are given
in table~\ref{tab-top-10} \cite{Re04}.  Also tabulated are the average age of
these citations and a measure we term citation impact; this is defined as the
product of the number of citations to a publication and the average age of
these citations.  This measure highlights publications that have influence
over a long time period.  The top 10 articles, according to this latter
measure, are listed in table~\ref{tab-rest}:

{\small\begin{longtable}{|c|p{0.3in}|>{\hfill}p{0.25in}|>{\hfill}p{0.375in}|%
>{\hfill}p{0.3in}|>{\hfill}p{0.35in}|>{\hfill}p{0.4in}|>{\hfill}p{0.43in}|p{1.8in}|p{2.0in}|}
\caption{Top-10 cited PR articles.  The asterisks denote citation undercount
due to citations with missing prepended A/B page numbers -- 123 out of 3227 total
for item 1 and 120 out of 2640 for item 2.}\label{tab-top-10}\\

\hline
  Impact & \multicolumn{4}{c|}{}            & \#~~    & Av. &        &       & \\
  Rank   & \multicolumn{4}{c|}{Publication} & cites & Age & Impact & Title & Author(s) \\ \hline
\endhead
  1& PR& 140& A1133& 1965&  3227*&   26.64&    85972& Self-Consistent Equations...
  & W. Kohn \& L. J. Sham \\ \hline
  2& PR& 136& B864& 1964&   2460*&   28.70&   70604 &  Inhomogeneous Electron Gas&
  P. Hohenberg \& W. Kohn \\ \hline
  3& PRB& 23& 5048& 1981&   2079&   14.38&   29896&     Self-Interaction Correction
  to...&  J. P. Perdew \& A. Zunger\\ \hline
 4& PRL& 45& 566& 1980&   1781&   15.42&   27463&      Ground State of the
  Electron ...& D. M. Ceperley \& B. J. Alder\\ \hline
 5& PR& 108& 1175& 1957&   1364&   20.18&   27526 &  Theory of
  Superconductivity&  \mbox{J. Bardeen, L. N. Cooper, \&} J.~R.~Schrieffer\\ \hline
 6& PRL& 19& 1264& 1967&   1306&   15.46&   20191 &   A Model of Leptons&
  S. Weinberg\\ \hline
 7& PRB& 12& 3060& 1975&   1259&   18.35&   23103 &   Linear Methods in Band
  Theory&  O. K. Andersen\\ \hline
  8& PR& 124& 1866& 1961&   1178&   27.97&   32949 &  Effects of Configuration...
  &  U. Fano\\ \hline
 8& RMP& 57& 287& 1985&   1055&   9.17&   9674&     Disordered Electronic
  Systems & P. A. Lee \& T. V. Ramakrishnan\\ \hline
 9& RMP& 54& 437& 1982&   1045&   10.82&   11307&    Electronic Properties of...
  & \mbox{T.~Ando, A.~B.~Fowler, \&} F.~Stern\\ \hline
 10& PRB& 13& 5188& 1976&   1023&   20.75&   21227&   Special Points for
  Brillouin-...&  H. J. Monkhorst \& J. D. Pack\\ \hline

\end{longtable}
}


{\small\begin{longtable}{|c|p{0.3in}|>{\hfill}p{0.25in}|>{\hfill}p{0.375in}|%
>{\hfill}p{0.3in}|>{\hfill}p{0.35in}|>{\hfill}p{0.4in}|>{\hfill}p{0.43in}|p{1.8in}|p{2.1in}|}
\caption{The top-10 PR articles ranked by citation impact.}\label{tab-rest}\\
\hline
  Cite & \multicolumn{4}{c|}{}            & \#~~    & Av. &        &       & \\
  Rank   & \multicolumn{4}{c|}{Publication} & cites & Age & Impact & Title & Author(s) \\ \hline
\endhead
  1& PR& 140& A1133& 1965&  3227*&   26.64&    85972& Self-Consistent Equations...
  & W. Kohn \& L. J. Sham \\ \hline
  2& PR& 136& B864& 1964&   2460*&   28.70&   70604 &  Inhomogeneous Electron Gas&
  P. Hohenberg \& W. Kohn \\ \hline
  3& PR& 124& 1866& 1961&   1178&   27.97&   32949 &  Effects of Configuration...
  &  U. Fano\\ \hline
  4& PR& 40& 749& 1932&   561&   55.76&   31281 &     On the Quantum Correction...
  & E. Wigner\\ \hline
  5& PRB& 23& 5048& 1981&   2079&   14.38&   29896&     Self-Interaction Correction
  to...&  J. P. Perdew \& A. Zunger\\ \hline
  6& PR& 82& 403& 1951&   643&   46.35&   29803&       Interaction Between 
  d-Shells ...& C. Zener\\ \hline
  7& PR& 47& 777& 1935&   492&   59.64&   29343 &     Can Quantum-Mechanical... &
  \mbox{A. Einstein, B. Podolsky, \&} N.~Rosen\\ \hline
  8& PR& 46& 1002& 1934&   557&   51.49&   28680 &    On the Interaction of... &  E. Wigner\\ \hline
  9& PR& 109& 1492& 1958&   871&   32.00&   27872&     Absence of Diffusion in...
  &  P. W. Anderson\\ \hline
 10& PR& 108& 1175& 1957&   1364&   20.18&   27526 &  Theory of
  Superconductivity&  \mbox{J. Bardeen, L. N. Cooper, \&} J.~R.~Schrieffer\\ \hline

\end{longtable}
}

\newpage

\section{Major Discoveries and Hot Papers}
\label{hot}

The following two tables provide the top-10 cited PR papers for which the
temporal history of citations allows one to characterize the publications
either as major discoveries (Table~\ref{tab-discovery}) or as hot
(Table~\ref{tab-hot}).  The articles in these tables are listed in
chronological order.

{\small\begin{longtable}{|>{\hfill}p{0.2in}|p{0.3in}|>{\hfill}p{0.25in}|>{\hfill}p{0.375in}|
>{\hfill}p{0.3in}|>{\hfill}p{0.35in}|>{\hfill}p{0.4in}|>{\hfill}p{0.43in}|p{1.9in}|p{1.9in}|}
\caption{Chronological list of the top-cited discovery papers with $>500$ 
citations and citation/paper age ratio $<0.4$.}\label{tab-discovery}\\ \hline
{ }  &  \multicolumn{4}{c|}{}             & \#~~    &Av.&& &\\ 
{ } &  \multicolumn{4}{c|}{Publication}  & cites & Age& Impact& Title & Author(s) \\ \hline
\endhead

  3&    PR& 125& 1067& 1962&   587&   7.02&   4120.7& Symmetries of Baryons
     \& Mesons & M. Gell-Mann\\ \hline
 14&    PR& 182& 1190& 1969&   563&   13.75&   7741.3& Nucleon-Nucleus
     Optical-Model Parameters, $A>40$, $E<50$ MeV   &\mbox{F. D. Becchetti, Jr. \&}
     G.~W.~Greenlees\\ \hline
 16&    PRD& 2& 1285& 1970&   738&   11.21&   8273.0& Weak Interactions with
     Lepton-Hadron Symmetry   &\mbox{S. L. Glashow, J. Iliopoulos, \&}
     L.~Maiani\\ \hline
 19&    PRD& 10& 2445& 1974&   577&   11.90&   6866.3& Confinement of Quarks &
     K. G. Wilson\\ \hline
 21&    PRL& 32& 438& 1974&   545&   11.14&   6071.3& Unity of All
     Elementary...& H. Georgi \& S. L. Glashow\\ \hline
 24&    PRD& 12& 147& 1975&   501&   10.66&   5340.7&  Hadron Masses in a
     Gauge Theory   &\mbox{A. De R\'ujula, H. Georgi, \&} S.~L.~Glashow\\ \hline
 27&    PRL& 53& 1951& 1984&   559&   7.89&   4410.5&  Metallic Phase with
     Long-Range Orientational Order and...&
     D. Shechtman, I. Blech, D. Gratias, \& J. W. Cahn\\ \hline
 28&    PRA& 33& 1141& 1986&   501&   6.44&   3226.4&  Fractal Measures and
     Their:...&T. C. Halsey et al.\\ \hline
 31&    PRL& 58& 2794& 1987&   525&   4.77&   2504.3&  Theory of high-$T_c$...& V. J. Emery\\ \hline
 33&    PRL& 58& 908& 1987&   625&   1.94&   1212.5&   Superconductivity at
     93K in...&
     M. K. Wu et al.\\ \hline
 38&    PRB& 43& 130& 1991&   677&   5.17&   3500.1&  Thermal Fluctuations,
     Quenched Disorder,...&\mbox{D. S. Fisher, M. P. A. Fisher, \&} D.~A.~Huse\\ \hline
 
\end{longtable}
}

{\small\begin{longtable}{|>{\hfill}p{0.3in}|>{\hfill}p{0.25in}|>{\hfill}p{0.375in}|
      >{\hfill}p{0.3in}|>{\hfill}p{0.35in}|>{\hfill}p{0.4in}|>{\hfill}p{0.43in}|p{1.9in}|p{2.0in}|}
\endhead
\caption{Chronological list of the 10 most cited ``hot'' papers (see text for
definition).}\label{tab-hot}\\ \hline

   \multicolumn{4}{|c|}{}             & \#~~    &Av.&& &\\ 
   \multicolumn{4}{|c|}{Publication}  & cites & Age& Impact& Title & Author(s) \\ \hline

PR& 40& 749& 1932&   561&   55.76&   31281 &     On the Quantum Correction... & E. Wigner\\ \hline
PR& 47& 777& 1935&   492&   59.64&   29343 &     Can Quantum-Mechanical\hfill\break
  Description of Physical...&  \mbox{A. Einstein, B. Podolsky, \&} N.~Rosen\\ \hline
  PR& 109& 1492& 1958&   871&   32.00&   27872&  Absence of Diffusion in...&P. W. Anderson\\ \hline
PR& 136& B864& 1964&   2460&   28.70&   70604 &  Inhomogeneous Electron Gas&  P. Hohenberg \& W. Kohn\\ \hline
 PR& 140& A1133 & 1965&  3227& 26.64&  85972&     Self-Consistent
Equations...& W. Kohn \& L. J. Sham\\ \hline
  PRB& 13& 5188& 1976&   1023&   20.75&   21227&  Special Points for Brillouin-...& H. J. Monkhorst \& J. D. Pack\\ \hline
  PRL& 48& 1425& 1982&   829&   15.05&   12477&  Efficacious Form for...&L. Kleinman \& D. M. Bylander\\ \hline
  PRB& 41& 7892& 1990&   691&   9.68&   6689&  Soft Self-Consistent Pseudo-...& D. Vanderbilt\\ \hline
  PRB& 45& 13244& 1992&   394&   8.08&   3184&  Accurate and Simple Analytic...& J. P. Perdew \& Y. Wang\\ \hline
  PRL& 70& 1895& 1993&   495&   7.36&   3643&  Teleporting an Unknown...&C. H. Bennett et al.\\ \hline

\end{longtable}
}

\newpage
\end{widetext}

\end{document}